\documentclass[prb,twocolumn,aps,preprintnumbers,amsmath,amssymb, showpacs]{revtex4}
\usepackage{graphicx}

\begin{document}
\preprint{}
\draft

\title{Loss-resistant state teleportation and entanglement swapping using a quantum-dot spin in an optical microcavity}

\author{C.Y.~Hu}\email{chengyong.hu@bristol.ac.uk}
\author{J.G.~Rarity}
\affiliation{Department of Electrical and Electronic Engineering, University of Bristol, University Walk,
Bristol BS8 1TR, United Kingdom}

\begin{abstract}

We present a scheme for efficient state teleportation and entanglement swapping using a single quantum-dot
spin in an optical microcavity based on giant circular birefringence.  State teleportation or entanglement
swapping is heralded by the sequential detection of two photons, and is finished
after the spin measurement. The spin-cavity unit works as a complete Bell-state analyzer
with a built-in spin memory allowing loss-resistant repeater operation.
This device can work in both the weak coupling and the strong coupling regime,
but high efficiencies and high fidelities are only achievable when the side leakage and cavity loss is low.
We assess the feasibility of this device, and show it can be implemented with current technology.
We also propose a spin manipulation method using single photons, which could be used to preserve the
spin coherence via spin echo techniques.

\end{abstract}

\date{\today}


\maketitle

\section{Introduction}

The key resource for quantum communication \cite{gisin02, bennett93, zukowski93} is entangled pairs (EPR pairs) of particles distributed over large distances.
Typically photons are used to encode and transport quantum bits (qubits) over long distance due to their relative insensitivity to decoherence. However the optical channels suffer from exponential loss and the resulting rate of generation of remote EPR pairs can be vanishingly small. In the last ten years, schemes for extending the range of quantum communication have been proposed \cite{cirac97, kimble08, briegel98, dur99, duan01, childress06, zhao07, loock06, munro08, waks02, jacobs02, collins05} using quantum teleportation, i.e., state teleportation and entanglement swapping. \cite{bennett93, zukowski93}
One crucial component for these schemes is efficient Bell-state analysis.
The Bell-state analyzer (BSA) makes a joint measurement on an incoming unknown qubit and one particle of an EPR pair and projects the two-qubit state onto one of the four Bell states. Knowing which Bell state they are in allows the state of the remote particle of the EPR pair to be rotated to match the state of the unknown qubit, thus performing state teleportation. If the unknown qubit is one partner in a remote EPR pair, the entanglement now extends between another two remote partner particles, which is entanglement swapping.
Simple but inefficient BSAs can be built using linear optics. For polarization-entangled photons, this
 consists of a 50/50 beam splitter and two polarizing beam splitters (PBS's) based on two-photon interference. \cite{hom87} This standard optical BSA  can identify only two of the four Bell states, so the corresponding teleportation is probabilistic (at most $50\%$ efficient).
\cite{bouwmeester97, riedmatten04, pan98} A further more damaging limitation is that this BSA relies on the synchronous arrival of two indistinguishable photons at the beam splitter. Both photons will have suffered loss and the probability for both arriving at the same time will be the product of these losses, which is typically much less than $50\%$.

In order to capture and link all photons that arrive at the BSA, a quantum memory is needed with some form of heralding to announce the photon arrival. In this way a photon arriving in one side of the system is stored until another arrives from the other side, at which point the two photons pass to a complete efficient BSA. Multi-link systems based on this approach can have a throughput of photons limited only by the losses of a single link independent of the number of links. However the fidelity of entanglement generated will be limited by the product of the fidelities of the individual links and would eventually drop below the threshold where the entanglement is useful. This can be remedied by integrating entanglement purification into a repeater as first introduced by Briegel et al. \cite{briegel98, dur99} to distribute entanglement with loss now a polynomial function of distance.

In this work, we present a device that incorporates a complete BSA with heralded single photon memory to provide a
loss-resistant repeater. The corresponding quantum teleportation is efficient (or deterministic in the ideal case),
heralded and loss resistant.  The device is based on a charged quantum-dot (QD) carrying a single spin coupled to an optical microcavity, and
exploits the photon-spin entangling gate that can be achieved in this system. \cite{hu08a, hu08b, hu09} This device can be extended to include
entanglement purification and we will report on this aspect elsewhere.

Compared with the standard optical BSAs based on two-photon interference,\cite{hom87} our BSA has three main advantages: (1) Indistinguishability
and synchronization of photons are not required as we exploit the spin coherence rather than the
photon coherence.  Quantum teleportation
is heralded by the sequential detection of two photons at different arrival time, and is finished after
the spin measurement.
(2) It measures all Bell states (is complete) and is loss-resistant. Both features, especially the latter,
can largely enhance the efficiency.
Our BSA has a built-in spin memory, so the time window for it to collect photons is determined by
the spin coherence time
which is in the ns or $\mu$s range, several orders of magnitude longer than the temporal overlap of photons (typically $<10~$ps) in standard
optical BSAs. As a result, the signal rate or the distance for quantum communication
can be significantly enhanced.
It is well known that a complete deterministic BSA using linear optics only is impossible. \cite{luetkenhaus99}
Although complete BSAs using optical \cite{kim01} or measurement-based \cite{klm01} nonlinearities are possible, they
often suffer from low efficiency.
Search for complete and efficient BSAs is always a big challenge in the field of quantum information
science. \cite{boschi98, kwiat98, walborn03b, schuck06, barbieri07, barreiro08, bonato10}
The efficiency of our  BSA
can be $100\%$ in the ideal case, and $>35\%$ in a realistic device with current technology.
We show in this work that this complete and efficient BSA could be implemented with current technology.
(3) It is versatile. Besides BSA, the spin-cavity unit can also work as various entangling gates,
photon-spin interface (spin memory), spin-controlled single photon source, and quantum non-demolition (QND) measurement. \cite{hu08a, hu08b, hu09}
The compatibility with standard semiconductor processing techniques allows all these
functions integrated onto a chip.

The paper is organized as follows: In Sec. II, we discuss the type-I BSA using a single QD-spin in a single-sided cavity with its application for state teleportation. In Sec. III, various experimental challenges in implementing this device are discussed. We calculate the BSA fidelity and efficiency in more realistic cavities. Results show that the BSA
can work in both the weak coupling and strong coupling regime, but high efficiencies and high fidelities are achieved
when the side leakage and cavity loss is low.  We also
introduce a spin manipulation method using single photons and its potential application to spin echo schemes to prolong the spin coherence time which limits the storage time of the intrinsic spin memory.
In Sec. IV, we show entanglement swapping using the type-I BSA.
In Sec. V, the type-II BSA using a single QD-spin in a double-sided cavity is presented with its applications
for state teleportation and entanglement swapping. Finally, we make our conclusions.

\section{Bell-state Analyzer (Type I)}

The optical selection rule of negatively or positively charged excitons (i.e., the trion $X^-$ or $X^+$) in QDs enhanced in a cavity-QED system
leads to large differences
in phase or amplitude of reflection/transmission coefficients between two circular polarisations of photons. This giant circular
birefringence (GCB) increases with increasing QD-cavity coupling strength and could be observed in both the strong and
the weak coupling regime. The GCB applications for quantum non-demolition (QND) measurement, various entangling gates, photon-spin interface,
and spin-controlled single photon source have been discussed in our previous work. \cite{hu08a, hu08b, hu09}

Here we consider a QD-confined electron spin in a single-sided microcavity (type-I) with the top mirror partially reflective and the bottom
mirror $100\%$ reflective.
In this spin-cavity system, GCB can manifest as the phase difference in the reflection coefficients
between $|R\rangle$ and $|L\rangle$ photons. By suitable detuning of the photon frequency, the phase difference can be adjusted to $\pm \pi/2$ and thus
a photon-spin entangling gate can be developed, which is described by a two-qubit phase shift operator \cite{hu08a, hu08b}
\begin{equation}
\hat{U}(\pi/2)=e^{i\pi/2(|L\rangle \langle L|\otimes |\uparrow \rangle\langle \uparrow| + |R\rangle\langle R|\otimes |\downarrow
\rangle\langle \downarrow|)},
\label{gate1}
\end{equation}
where $|R\rangle$, $|L\rangle$ are right-circular and left-circular photon polarization states, and $|\uparrow \rangle$, $|\downarrow \rangle$
are the spin eigenstates along the optical axis, i.e, photon input/output direction.
If the side-leakage and loss rate $\kappa_s$ is lower than the output coupling rate $\kappa$, near unity reflectance can be achieved for the empty cavity
at all frequencies and for the trion-coupled cavity around the central frequency region in the strong coupling regime, \cite{hu08a, hu08b}
so the gate is deterministic in this ideal case and we will discuss the details in Sec. III.

Note that the two-qubit phase gate is different from the controlled-Z (CZ) gate, i.e., $\hat{U}(\pi/2)\neq \hat{U}_{CZ}$. This can be
easily seen by checking their matrix representations, which are
\begin{equation}
\hat{U}(\pi/2)=\begin{pmatrix}
1 & 0 & 0 & 0 \\
0 & i & 0 & 0 \\
0 & 0 & i & 0 \\
0 & 0 & 0 & 1 \\
\end{pmatrix}
\quad \hat{U}_{CZ}=\begin{pmatrix}
1 & 0 & 0 & 0 \\
0 & 1 & 0 & 0 \\
0 & 0 & 1 & 0 \\
0 & 0 & 0 & -1 \\
\end{pmatrix}.
\end{equation}
For our gate, if the input state is $|0\rangle|1\rangle$ or $|1\rangle|0\rangle$, it induces a phase shift of $\pi/2$ on the output state; if the input state is $|0\rangle|0\rangle$ or $|1\rangle|1\rangle$, it has no effect on the output
states. For the controlled-Z gate, only if the input state is $|1\rangle|1\rangle$, it induces a phase shift of $\pi$ on  the output state, but has no influence otherwise. Due to the side leakage and the cavity loss, it is hard
to achieve a phase shift of $\pi$ in a realistic cavity-QED system. \cite{duan04} However, the condition for the $\pi/2$ phase shift is actually achievable in a realistic system with losses
as discussed in Sec. III. It is worthy to point out that our phase gate exploits the coherent
photon-spin interaction in the linear region, therefore it is very different from other quantum phase
gates based on Kerr nonlinearity in the nonlinear region. \cite{turchette95, imamoglu97, gheri98}

\begin{figure}[ht]
\centering
\includegraphics* [bb= 140 400 489 678, clip, width=5.8cm, height=5cm]{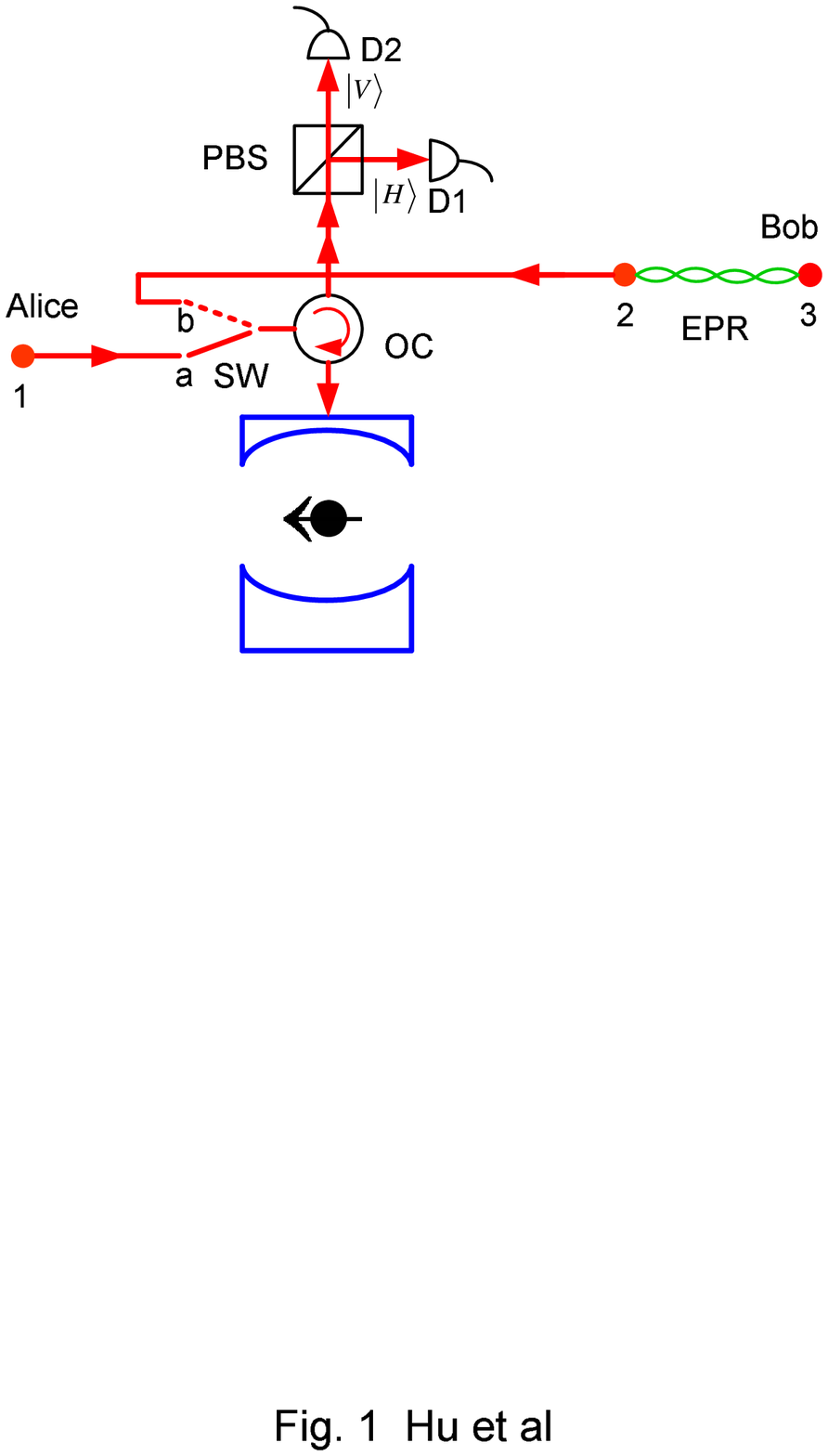}
\caption{(color online). Schematic of state teleportation with a QD-spin in a single-sided microcavity (type-I).
In this case, the spin-cavity unit works as a three-photon GHZ-state generator and a complete Bell-state analyzer.
SW (optical switch), OC (optical circulator), PBS (polarizing beam splitter), D1 and D2 (photon detectors).} \label{fig1}
\end{figure}

First we show state teleportation with the spin-cavity unit (see Fig. 1). Suppose Alice wants to transfer photon 1 in a unknown
state $|\psi^{ph}\rangle_1=\alpha|R\rangle_1+\beta|L\rangle_1$ to  Bob and Alice holds the spin-cavity unit. Alice and Bob share a pair
of entangled photons 2 and 3 in the state $|\psi^{ph}\rangle_{23}=(|R\rangle_2|L\rangle_3 + |L\rangle_2|R\rangle_3)/\sqrt{2}$ (photon 2
to Alice and photon 3 to Bob).  The EPR pair can be generated either by the spontaneous parametric down conversion process, \cite{gisin02} or
the photon entangler, \cite{hu08b} or the entanglement beam splitter.\cite{hu09} The electron spin is initialized
to $|\psi^s\rangle=(|\uparrow\rangle+|\downarrow\rangle)/\sqrt{2}$. Given the losses the arrival of photon 1 should be heralded before
photon 2. Initially an optical switch directs photon 1 from Alice to the system until it is detected in D1 or D2 as shown in Fig. 1. The switch
is then switched to await the detection of photon 2. The time difference between photons 1 and 2 should be less than the spin coherence
 time $T^e_2$. After the reflection of photons 1 and 2,
 the total state of three photons with one spin is transformed into
\begin{equation}
\begin{split}
|\psi&^{ph}\rangle_1 \otimes |\psi^{ph}\rangle_{23} \otimes |\psi^s\rangle \xrightarrow[\text{~to photon 1,2}]{\hat{U}(\pi/2)} \\
\frac{1}{\sqrt{2}}\{&[\alpha |R\rangle_1|R\rangle_2|L\rangle_3-\beta |L\rangle_1|L\rangle_2|R\rangle_3]|-\rangle\\
+i&[\alpha|R\rangle_1|L\rangle_2|R\rangle_3+\beta|L\rangle_1|R\rangle_2|L\rangle_3]|+\rangle\},
\end{split}
\label{tele1}
\end{equation}
where $|\pm\rangle=(|\uparrow\rangle\pm|\downarrow\rangle)/\sqrt{2}$.

If Alice measured the spin in the $|+\rangle$, $|-\rangle$ basis (e.g., by spin rotations and optical QND measurement \cite{hu08a, hu08b}),
the total state would be projected to two partially entangled GHZ states \cite{greenberger90} which contain the state information Alice wants
to transfer. We see here the spin-cavity unit works as a three-photon GHZ state generator. The spin measurement is not carried out until the
detection of photons 1 and 2.  By expressing photons 1 and 2 in the $|H\rangle$, $|V\rangle$ basis with a PBS, the
right side of Eq. (\ref{tele1}) becomes
\begin{equation}
\begin{split}
\frac{1}{2\sqrt{2}}\{&[|H\rangle_1|H\rangle_2-|V\rangle_1|V\rangle_2]|-\rangle(\alpha|L\rangle_3-\beta|R\rangle_3)\\
+i&[|H\rangle_1|V\rangle_2+|V\rangle_1|H\rangle_2]|-\rangle(\alpha|L\rangle_3+\beta|R\rangle_3)\\
+i&[|H\rangle_1|H\rangle_2+|V\rangle_1|V\rangle_2]|+\rangle(\alpha|R\rangle_3+\beta|L\rangle_3)\\
+&[|H\rangle_1|V\rangle_2-|V\rangle_1|H\rangle_2]|+\rangle(\alpha|R\rangle_3-\beta|L\rangle_3)\}.
\end{split}
\label{tele2}
\end{equation}
After the photon polarization and the spin measurements, photon 3 at Bob's hand will be found in a state, which
is related to the initial state and unequivocally associated to the measurement results consisting of polarizations of photons 1 and 2
in the $|H\rangle$, $|V\rangle$ basis and the spin in the $|+\rangle$, $|-\rangle$ basis (see Table \ref{tab1}). By applying the appropriate single-qubit
gate \cite{bennett93} on photon 3, Bob gets the initial state Alice wants to transfer. At this point the teleportation procedure is
complete.

\begin{table}[ht]
\caption{The correspondence between the photon 1, 2 polarization and the spin measurement results and the photon 3 states.}
\begin{ruledtabular}
\begin{tabular}{lll}
\textrm{Photons 1, 2}&
\textrm{Spin} &
\textrm{Photon 3}\\
\colrule
$|H\rangle_1|H\rangle_2$ or $|V\rangle_1|V\rangle_2$ & $|-\rangle$ & $\alpha|L\rangle_3-\beta|R\rangle_3$ \\
$|H\rangle_1|V\rangle_2$ or $|V\rangle_1|H\rangle_2$ & $|-\rangle$ & $\alpha|L\rangle_3+\beta|R\rangle_3$ \\
$|H\rangle_1|H\rangle_2$ or $|V\rangle_1|V\rangle_2$ & $|+\rangle$ & $\alpha|R\rangle_3+\beta|L\rangle_3$ \\
$|H\rangle_1|V\rangle_2$ or $|V\rangle_1|H\rangle_2$ & $|+\rangle$ & $\alpha|R\rangle_3-\beta|L\rangle_3$ \\
\end{tabular}
\end{ruledtabular}
\label{tab1}
\end{table}

To investigate the role of the spin-cavity unit further, we consider the two-photon Bell states as the input, i.e.,
$|\Psi^{\pm}\rangle=(|R\rangle_1|L\rangle_2 \pm |L\rangle_1|R\rangle_2)/\sqrt{2}$ and
$|\Phi^{\pm}\rangle=(|R\rangle_1|R\rangle_2 \pm |L\rangle_1|L\rangle_2)/\sqrt{2}$, and get the following transformations
\begin{equation}
\begin{split}
&\hat{U}(\pi/2)|\Psi^{\pm}\rangle|+\rangle=i|\Psi^{\pm}\rangle|+\rangle \\
&\hat{U}(\pi/2)|\Phi^{\pm}\rangle|+\rangle=|\Phi^{\mp}\rangle|-\rangle.
\end{split}
\label{bell1}
\end{equation}
Obviously, the spin measurements in the $|+\rangle$, $|-\rangle$ basis can distinguish $|\Psi^{\pm}\rangle$ from  $|\Phi^{\pm}\rangle$,
and the photon polarization measurement in the $|H\rangle$, $|V\rangle$ basis can distinguish between $|\Psi^{+}\rangle$ and $|\Psi^{-}\rangle$
and between $|\Phi^{+}\rangle$ and $|\Phi^{-}\rangle$. So the spin-cavity unit with a PBS is
a complete BSA.
If expressing photons 1 and 2 in the $|\Psi^{\pm}\rangle$, $|\Phi^{\pm}\rangle$ basis on the right side of Eq. (\ref{tele1}),
we get the same result as Eq. (\ref{tele2}).

As the spin-cavity unit also works as a photon-spin interface, \cite{hu08b} our BSA contains a built-in spin memory with the storage
time limited by the spin coherence time $T^e_2$.
The above teleportation procedure can be understood in another way: On detecting photon 1, the initial state is transferred to the spin
and stored on it; On detecting photon 2, the spin
and photon 3 get entangled; After the spin measurement, the state is transferred from the spin to photon 3.
As we exploit coherent photon-spin interaction rather than the two-photon interference,
our BSA does not require synchronized and indistinguishable photons in contrast to the standard
optical BSA \cite{bouwmeester97, riedmatten04, pan98} The synchronization should be maintained
within the
photon coherence time (typically $<10~$ps) for the standard optical BSA, whereas the time window
for our BSA to capture photons is equal to the spin coherence time $T^e_2$.
As discussed in Sec. III, the electron spin decoherence time is normally several ns for InAs- or GaAs-based
QD when no spin protection methods are applied. In this case, the signal rate for quantum communication
using our BSA is enhanced by up to two orders of magnitude,\cite{rate} and  the distance at the same signal rate
(or the length of each repeater segment) is increased by up to 5 times,
compared with that using the standard optical BSA.
Using the spin echo techniques (see Sec. III), $T^e_2$ could be extended to the $\mu$s range. As
a result, the signal rate could be enhanced by up to five orders of magnitude, or
 the distance by 12 times. Therefore,
our BSA can combat losses in quantum channels, and holds great potential for
long-distance quantum communication, \cite{schmitt07, fedrizzi09, jin10}
For example, the longest distance for quantum communication is currently kept around
$100~$km. With our BSA, the distance (without the help of quantum reptears) could be extended
to over $1000~$km at which most satellites are covered. \cite{aspelmeyer03b}

\section{Experimental challenges}

In this Section, we discuss the feasibility to implement the BSA function in a promising system
with GaAs- or InAs-based QDs in micropillar microcavities. This type of microcavity with circular
cross section can support circularly polarized light and have negligible mode mismatching between the traveling
and the cavity photons. \cite{rakher09}

To calculate the fidelity and efficiency for the entanglement analysis, we have to use the reflection operator, rather
than the phase shift operator $\hat{U}(\pi/2)$,  to describe the photon-spin entangling
gate, \cite{hu08a, hu08b} i.e.,
\begin{equation}
\begin{split}
\hat{r}(\omega)= & r_0(\omega)(|R\rangle \langle R|\otimes |\uparrow \rangle\langle \uparrow| + |L\rangle\langle L|\otimes |\downarrow
\rangle\langle \downarrow|) \\
+ & r_h(\omega)(|L\rangle \langle L|\otimes |\uparrow \rangle\langle \uparrow| + |R\rangle\langle R|\otimes |\downarrow
\rangle\langle \downarrow|),
\end{split}
\label{gate2}
\end{equation}
where $r_0(\omega)\equiv|r_0(\omega)|e^{i\varphi_0(\omega)}$ and $r_h(\omega)\equiv|r_h(\omega)|e^{i\varphi_h(\omega)}$ are the reflection coefficients
for the empty (or cold)  cavity (with trion uncoupled to the cavity) and hot cavity (with trion coupled to the cavity), respectively. In the weak excitation approximation where the real excitation can be neglected, \cite{note0}  $r_h(\omega)$ and $r_0(\omega)$
are given by \cite{hu08a}
\begin{equation}
\begin{split}
&r_h(\omega)=1-\frac{\kappa[i(\omega_{X^-}-\omega)+\frac{\gamma}{2}]}{[i(\omega_{X^-}-\omega)+
\frac{\gamma}{2}][i(\omega_c-\omega)+\frac{\kappa}{2}+\frac{\kappa_s}{2}]+\text{g}^2}\\
&r_0(\omega)=\frac{i(\omega_c-\omega)-\frac{\kappa}{2}+\frac{\kappa_s}{2}}{i(\omega_c-\omega)+\frac{\kappa}{2}+\frac{\kappa_s}{2}},\\
\end{split}
\label{reflec1}
\end{equation}
where $\omega$, $\omega_c$, $\omega_{X^-}$ are the frequencies of
input photon, cavity mode, and $X^-$ transition, respectively. g is the coupling strength between $X^-$ and the cavity mode.
$\gamma/2$ is the $X^-$ dipole decay rate, and $\kappa/2$, $\kappa_s/2$ are the cavity field decay rate into the output modes,
and the leaky modes (e.g., side leakage and cavity loss), respectively. The GCB effect comes from  $r_0(\omega)\neq r_h(\omega)$ in either the phase
or the reflectance.
Instead of the dispersive interaction, \cite{raimond01} we consider the resonant interaction with $\omega_c=\omega_{X^-}=\omega_0$ in this work.
Note that the reflection operator in Eq. (\ref{gate2}) consists of the contributions from both the empty and the hot cavity, and is
is the general and rigorous form from which we derive the phase shift operator $\hat{U}(\pi/2)$.

\begin{figure}[ht]
\centering
\includegraphics* [bb= 76 249 533 746, clip, width=7.5cm, height=7.5cm]{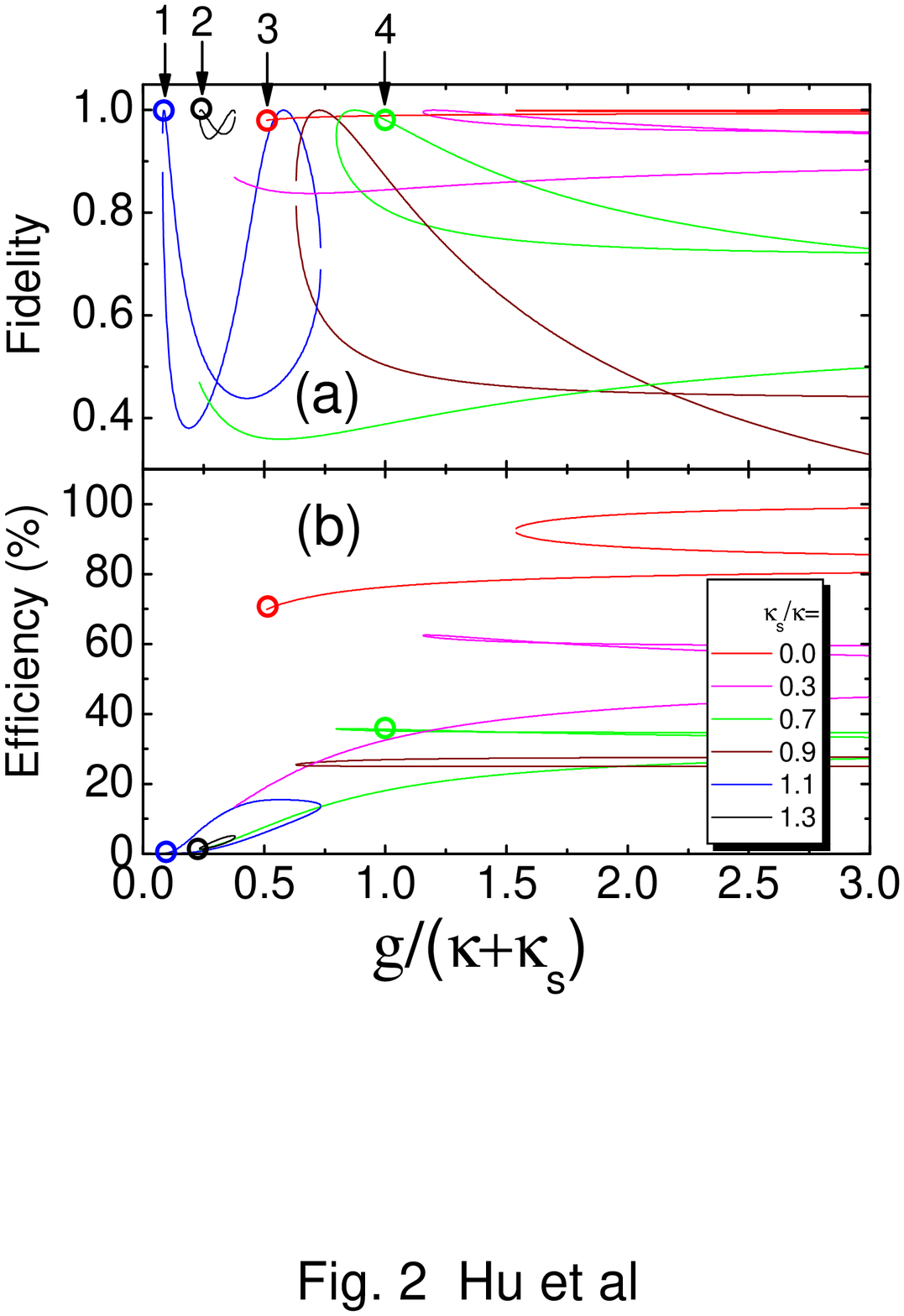}
\caption{(color online). Calculated results: (a) the fidelity $F^{(\Phi^{\pm})}$ and (b) the efficiency $\eta$ for the type-I BSA  as a function of the  normalized
coupling strength $\text{g}/(\kappa+\kappa_s)$ for different  $\kappa_s/\kappa$ values. Note that $F^{(\Psi^{\pm})}=1$ (not shown here) and $\eta=\eta^{(\Psi^{\pm})}=\eta^{(\Phi^{\pm})}$.} \label{fig2}
\end{figure}

If the side-leakage and cavity loss rate $\kappa_s$ is much lower than the output coupling rate $\kappa$ (the ideal case), we have $|r_0(\omega)|\simeq 1$ at all  $\omega$'s for the empty cavity, and $|r_h(\omega)|\simeq 1$ around the central frequency regime for the hot cavity in the strong coupling regime. The reflection operator $\hat{r}(\omega)$ in Eq. (\ref{gate2}) can be simplified as the phase shift operator $\hat{U}(\pi/2)$ in Eq. (\ref{gate1}) with unity fidelity and efficiency. However, when the side leakage and cavity loss
can not be neglected, the empty and the hot cavity have different reflectance in general, and the gate fidelity and efficiency
will be reduced. The fidelity (in amplitude) is given by
\begin{equation}
\begin{split}
& F^{(\Psi^{\pm})}=1 \\
& F^{(\Phi^{\pm})}=\frac{1}{\sqrt{1+\frac{1}{4}\left[\frac{|r_0(\omega^{\prime})|}
{|r_h(\omega^{\prime})|}-\frac{|r_h(\omega^{\prime})|}{|r_0(\omega^{\prime})|}\right]^2}}
\end{split}
\label{fid1}
\end{equation}
and the efficiency is
\begin{equation}
\eta=\eta^{(\Psi^{\pm})}=\eta^{(\Phi^{\pm})}=\frac{1}{4}[|r_0(\omega^{\prime})|^2+|r_h(\omega^{\prime})|^2]^2,
\end{equation}
where $\omega^{\prime}$ satisfies $\varphi_h(\omega^{\prime})-\varphi_0(\omega^{\prime})=\pm \pi/2$. The fidelity
to analyze $|\Psi^{\pm}\rangle$ remains unity, whereas the fidelity to analyze $|\Phi^{\pm}\rangle$  is generally less than one,
depending on how different $|r_0(\omega^{\prime})|$ and $|r_h(\omega^{\prime})|$ are.
Fig. 2 presents the numerical calculations of the BSA fidelity and efficiency by taking the trion decay rate $\gamma$ several $\mu$eV which is common
for InAs- or GaAs-based QDs, and much smaller than the cavity decay rate of interest.
The condition of $\varphi_h(\omega^{\prime})-\varphi_0(\omega^{\prime})=\pm \pi/2$ can be satisfied at various frequency detunings
leading to multiple fidelity and efficiency curves for each loss ratio $\kappa_s/\kappa$. \cite{note1}
If $\kappa_s > 1.3\kappa$, the phase shift $\pm \pi/2$ can not be achieved, and
the above discussions are invalid, i.e., the proposed BSA does not work.
If $\kappa_s <1.3\kappa$, a phase shift of $\pm \pi/2$ can be achieved, and
the BSA can work
in both the strong coupling regime with $\text{g}>(\kappa+\kappa_s)/4$ and the weak coupling regime with $\text{g}<(\kappa+\kappa_s)/4$.
We illustrate this by highlighting points 1-4 in Fig. 2. In the weak coupling regime, we can achieve unity fidelity, i.e., $F^{(\Phi^{\pm})}=F^{(\Psi^{\pm})}=1.0$ at points 1 and 2. However the efficiency is quite low: $\eta=0.03\%$ at point 1
where $\text{g}/(\kappa+\kappa_s)=0.086$  and $\kappa_s/\kappa=1.1$, and $\eta=1.4\%$ at point 2
where $\text{g}/(\kappa+\kappa_s)=0.24$  and $\kappa_s/\kappa=1.3$. In the strong coupling regime,
both the fidelity and the efficiency can be high. In the ideal case where
$\text{g}/(\kappa+\kappa_s)>1.5$ and $\kappa_s/\kappa \ll 1$, near unity fidelity and
near unity efficiency can be simultaneously achieved, and the reflection operator $\hat{r}(\omega)$
in Eq. (\ref{gate2}) can be simplified as the phase shift operator $\hat{U}(\pi/2)$ in Eq. (\ref{gate1}).
This case is exactly what we discussed in our previous work. \cite{hu08a, hu08b}
Point 3 represents the small loss limit ($\kappa_s/\kappa \approx 0$) where
we can achieve near unity fidelity ($F^{(\Phi^{\pm})}=0.98$, $F^{(\Psi^{\pm})}=1.0$) and efficiency
$\eta=69.9\%$ at $\text{g}/(\kappa+\kappa_s)=0.51$. Also we include point 4 which is experimentally achievable
in the strong coupling regime (see the following arguments). At point 4, near unity fidelity ($F^{(\Phi^{\pm})}=0.98$, $F^{(\Psi^{\pm})}=1.0$)
is associated with $\eta=35.6\%$ efficiency when $\text{g}/(\kappa+\kappa_s)\simeq 1.0$ and $\kappa_s/\kappa \simeq 0.7$.

It is easy to achieve the weak coupling experimentally, however, the strong coupling which is more challenging
has also been observed in various QD-cavity
systems \cite{reithmaier04, yoshie04, peter05, reitzenstein07, note2} and
 $\text{g}/(\kappa+\kappa_s) \simeq 0.5$ was reported \cite{reithmaier04} for $d=1.5~\mu$m micropillar microcavities
with a quality factor $Q=8800$.  By improving the sample designs and growth, \cite{reitzenstein07} the quality factors
for the micropillars of the same size were
increased to $\sim 4\times 10^4$, corresponding to $\text{g}/(\kappa+\kappa_s)\simeq 2.4$ (Ref. \onlinecite{note3}).
However, the quality factors in these micropillars are dominated by the side leakage and cavity loss rate, rather than the
output coupling rate,\cite{reitzenstein07} i.e., $\kappa_s/\kappa\gg 1$,  which is not what we want.
In order to reduce $\kappa_s/\kappa$, we could take such high-Q micropillars
and thin down the top mirrors to decrease the quality factor to  $Q\simeq 1.7\times 10^4$.
This process increases $\kappa$, whereas keeps $\kappa_s$ nearly unchanged (or slightly reduced).
As a result, we get $\kappa_s/\kappa \simeq 0.7$, whereas
the system now with $\text{g}/(\kappa+\kappa_s)\simeq 1.0$ remains in the strong coupling regime, corresponding to the point 4 in  Fig. 2.
Therefore, the proposed BSA and all related schemes could be implemented with current technology.
Small $\kappa_s/\kappa$ in the strong coupling regime is highly demanded for high efficiency operation and could be quite challenging for micropillar microcavities.

Due to the spin decoherence, the fidelity in Eq. (\ref{fid1}) decreases by a factor $F^{\prime}$,
\begin{equation}
F^{\prime}=[1+\exp{(-\Delta t/T^e_2)}]/2,
\label{fid2}
\end{equation}
where $T^e_2$ is the electron spin coherence time and $\Delta t$ is the time interval between two input photons for Bell state analysis.
To get high fidelity, the time interval between two photons should be shorter than
the spin coherence time $T^e_2$, i.e., $\Delta t<T^e_2$. To make the weak excitation
approximation valid, $\Delta t$ should be longer than $\tau/n_0 \sim$ns, \cite{note4} where $\tau$ is the
cavity photon lifetime and $n_0$ is the critical photon number of the spin-cavity system. \cite{kimble94}
As discussed later, $T^e_2$ could be extended to  $~\mu$s using the
spin echo techniques.

The trion dephasing can also reduce the  fidelity by a factor of $\tau/T_2$, where
$\tau$ is the cavity photon lifetime and $T_2$ is the trion coherence time. Two kinds of dephasing
processes should be considered here: the optical dephasing and the spin dephasing of $X^-$. It is well-known that the optical coherence time
of excitons in self-assembled In(Ga)As QDs can be as long as several hundred picoseconds, \cite{borri01, birkedal01, langbein04}
which is ten times longer than
the cavity photon lifetime (around tens of picoseconds in the strong coupling regime for cavity Q-factor of $10^4-10^5$). So the optical
dephasing can only slightly reduce the  fidelity by a few percent. The spin dephasing
of the $X^-$ is mainly due to the hole-spin dephasing. In the absence of the contact hyperfine interaction that happens for
holes, \cite{heiss07, gerardot08}
the QD-hole spin is expected to have long coherence time, and $T^h_2>100~$ns has been reported recently. \cite{brunner09} This spin
coherence time is at least three orders of magnitude longer than the cavity photon lifetime, so the spin dephasing of the $X^-$
can be safely neglected in our considerations.

For a realistic QD, the optical selection rule is not perfect due to the heavy-light hole mixing. This can reduce the
fidelity by a few percent as the hole mixing in the valence band is in the order of a few percent \cite{bester03}
[e.g., for self-assembled In(Ga)As QDs]. The hole mixing  could  be reduced by engineering the shape and size of
QDs or choosing different types of QDs. Note that our schemes are immune to the fine structure splitting
as it occurs for neutral excitons, but not for charged excitons due to the quenched exchange
interaction, \cite{bayer02, finley02} which is in accordance with Kramer's theorem.

Recently, significant progress has been made on optical spin cooling \cite{atature06, xu07} and optical spin manipulating \cite{gupta01, berezovsky08, press08, greilich09} in QDs. For our schemes, we could alternatively apply the
phase gate $\hat{U}(\pi/2)$ to perform single-shot QND measurement and initialize or read out the spin states  via single-photon
measurement. \cite{hu08a, hu08b, hu09}
Spin manipulation is also possible by using the
phase gate $\hat{U}(\pi/2)$. One photon in the $|R\rangle$ or $|L\rangle$ state can make
$90^{\circ}$ spin rotations around the optical axis
after the photon is reflected from the cavity with the transformation
\begin{equation}
\begin{split}
& \hat{U}(\pi/2)|R\rangle(\alpha|\uparrow\rangle+\beta|\downarrow\rangle)=|R\rangle
(\alpha|\uparrow\rangle+i\beta|\downarrow\rangle)\\
& \hat{U}(\pi/2)|L\rangle(\alpha|\uparrow\rangle+\beta|\downarrow\rangle)=|L\rangle
(i\alpha|\uparrow\rangle+\beta|\downarrow\rangle).
\end{split}
\end{equation}
In our previous work on this subject, \cite{hu08a} we have discussed the giant optical Faraday rotations
due to a single spin. Here is the reverse process where the giant spin rotations are induced by a single photon
(interacting with the spin in the cavity). The speed for this spin rotation is determined
by the photon coherence time, which can be several tens of ps. \cite{note4}
Similarly, two  photons in the $|R\rangle_1|R\rangle_2$
or $|L\rangle_1|L\rangle_2$ states can induce $180^{\circ}$ spin rotations after their reflections from the cavity
with the transformation
\begin{equation}
\begin{split}
& \hat{U}(\pi/2)|R\rangle_1|R\rangle_2(\alpha|\uparrow\rangle+\beta|\downarrow\rangle)=|R\rangle_1|R\rangle_2
(\alpha|\uparrow\rangle-\beta|\downarrow\rangle)\\
& \hat{U}(\pi/2)|L\rangle_1|L\rangle_2(\alpha|\uparrow\rangle+\beta|\downarrow\rangle)=|L\rangle_1|L\rangle_2
(-\alpha|\uparrow\rangle+\beta|\downarrow\rangle).
\end{split}
\label{bell2}
\end{equation}
The speed for this spin rotation is determined
by the time interval of two photons, which can be in the ns range. \cite{note4}
Spin rotations with other angles are also possible if we tune the photon frequency to get different phase
shift $\Delta\varphi$. The state transformation can be written as
\begin{equation}
\begin{split}
&\hat{U}(\Delta \varphi)|R\rangle(\alpha|\uparrow\rangle+\beta|\downarrow\rangle)=
|R\rangle[\alpha|\uparrow\rangle+e^{i\Delta \varphi}\beta |\downarrow\rangle]\\
&\hat{U}(\Delta \varphi)|L\rangle(\alpha|\uparrow\rangle+\beta|\downarrow\rangle)=
|L\rangle[e^{i\Delta \varphi}\alpha |\uparrow\rangle+\beta|\downarrow\rangle],
\end{split}
\label{bell3}
\end{equation}
and $+\Delta\varphi$ or $-\Delta\varphi$ spin rotations could be performed.
N photons in sequence (or N passes of a single photon) in the $|R\rangle_1|R\rangle_2\cdot\cdot\cdot|R\rangle_N$
or $|L\rangle_1|L\rangle_2\cdot\cdot\cdot|L\rangle_N$ states could be sent to the cavity in sequence
and make $+N\Delta\varphi$ or $-N\Delta\varphi$ spin rotations.
The phase shift $\Delta\varphi$ is tuneable between
$-\pi$ and $+\pi$ by tuning the photon frequency in an ideal system. The range of $\Delta\varphi$
is reduced by the cavity
side leakage and the cavity loss as discussed in our previous work. \cite{hu08a}
This spin manipulation method based on single photons is compatible with our schemes, and is
different from another optical method based on ac Stark effect
as demonstrated recently. \cite{gupta01, berezovsky08, press08, greilich09}

As discussed above, spin coherence time $T^e_2$ is important as our schemes rely on the coherent photon-spin interaction: (1) The storage time
of the spin memory is limited by $T^e_2$; (2) The entanglement fidelity for the BSA depends on
$T^e_2$ by Eq.(\ref{fid2});
(3) The time window for the BSA to capture photons is determined by $T^e_2$. The longer the $T^e_2$,
the bigger the time window and the stronger the signal rate (or more loss can be resisted).
The electron spin coherence time in GaAs-based or InAs-based QD is normally quite short ($T^{e}_2 \sim $ns) \cite{petta05, koppens08}
due to the hyperfine interaction
between the electron spin and $10^4$ to $10^5$ host nuclear spins, whereas the electron spin relaxation time is much longer ($T^e_1\sim$ms) \cite{elzerman04, kroutvar04}
due to the suppressed electron-phonon and spin-orbit interactions in QDs.  If nuclear spin fluctuations are totally suppressed,
$T^e_2$ will be prolonged and is limited by  $T^e_2 \leq 2T^e_1$.
Isotope engineering techniques are not suitable here as elements In, Ga and As do not have stable isotopes with zero-spin nuclei.
Dynamical decoupling based on spin echo techniques have been demonstrated to suppress the nuclear spin fluctuations effectively
and prolong the electron spin coherence $T^e_2$ to the $\mu$s range. \cite{petta05, koppens08, greilich09, clark09, press10}
However, our schemes are not compatible with such ESR-based spin manipulations in an external magnetic field.  As discussed above, two single photons can play the role of the $\pi$-pulse to
make the $180^{\circ}$ spin rotation around the optical axis.
Therefore we could still apply the spin echo techniques to
protect the spin coherence with various pulse sequences,
\cite{hahn50, carr54, meiboom58, khodjasteh05, yao07, witzel07, uhrig07} but now using single
photon pulses, rather than microwave pulses \cite{petta05, koppens08} or optical pulses. \cite{greilich09, clark09, press10}

\section{Entanglement swapping using type-I Bell-state analyzer}

\begin{figure}[ht]
\centering
\includegraphics* [bb= 22 400 520 678, clip, width=7cm, height=4.5cm]{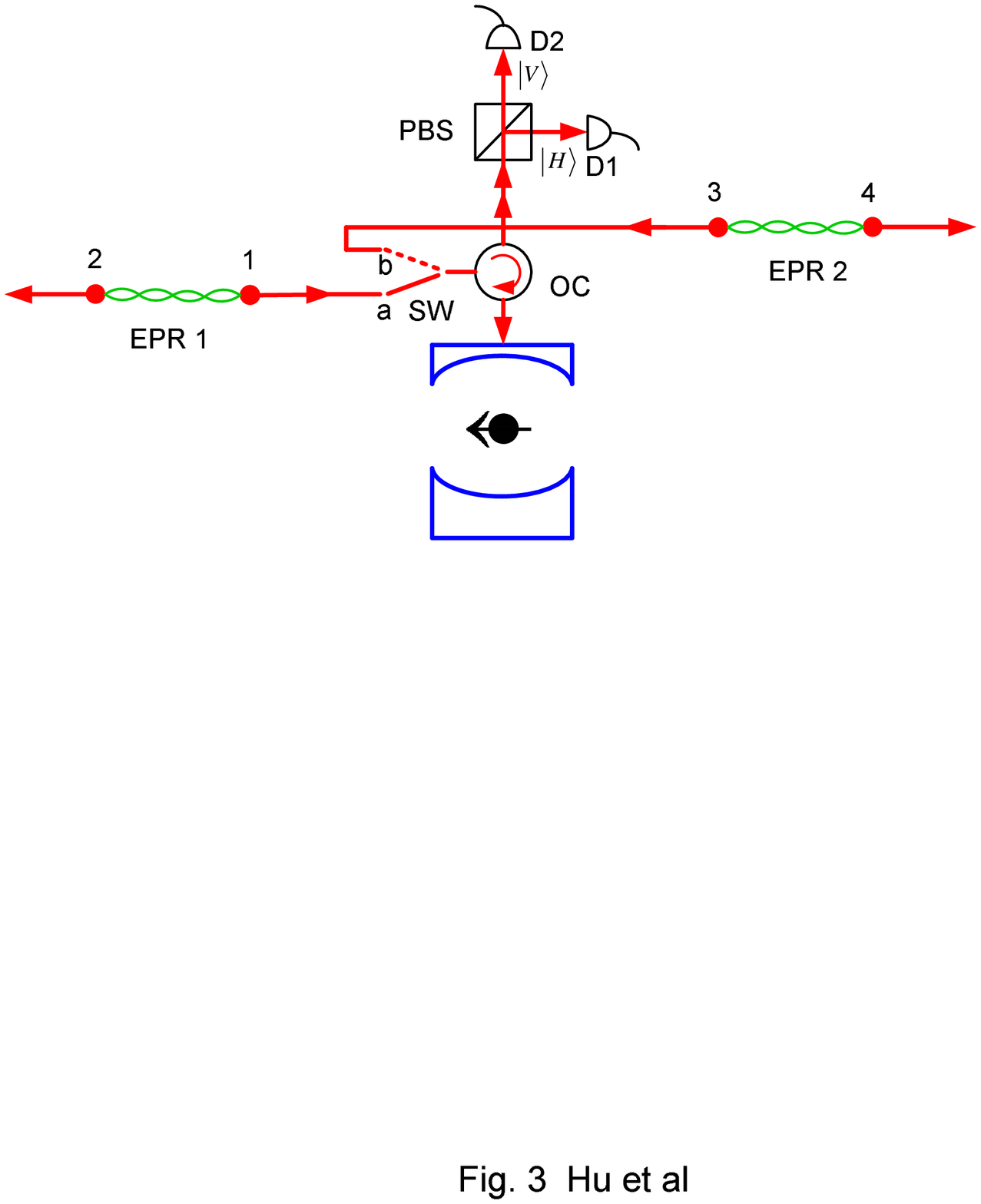}
\caption{(color online). Schematic of entanglement swapping between two EPR pairs with the type-I spin-cavity unit.
In this case, the spin-cavity unit works as
a four-photon GHZ-state generator and a complete Bell-state analyzer.} \label{fig3}
\end{figure}

As well as state teleportation, the BSA is also the key component for entanglement swapping, dense coding, and fault tolerant
quantum computing. As an example, we perform entanglement swapping with the spin-cavity unit (see Fig. 2). We prepare two
independent EPR pairs:  photon 1,2 in
the state $|\psi^{ph}\rangle_{12}=(|R\rangle_1|L\rangle_2+|L\rangle_1|R\rangle_2)/\sqrt{2}$, and photon 3,4 in the
state $|\psi^{ph}\rangle_{34}=(|R\rangle_3|L\rangle_4+|L\rangle_3|R\rangle_4)/\sqrt{2}$.
The spin is initialized to $|\psi^s\rangle=|+\rangle$. Photons 1 and 3 are then targeted to the spin-cavity unit.
After the reflection of photons 1 and 3, the total state of four photons with one spin is transformed into
\begin{equation}
\begin{split}
& |\psi^{ph}\rangle_{12} \otimes |\psi^{ph}\rangle_{34} \otimes |\psi^{s}\rangle \xrightarrow[\text{~to photon 1,3}]{\hat{U}(\pi/2)} \\
\frac{1}{4}\{&[|H\rangle_1|H\rangle_3-|V\rangle_1|V\rangle_3]|-\rangle[|L\rangle_2|L\rangle_4-|R\rangle_2|R\rangle_4]\\
+i&[|H\rangle_1|V\rangle_3+|V\rangle_1|H\rangle_3]|-\rangle[|L\rangle_2|L\rangle_4+|R\rangle_2|R\rangle_4]\\
+i&[|H\rangle_1|H\rangle_3+|V\rangle_1|V\rangle_3]|+\rangle[|L\rangle_2|R\rangle_4+|R\rangle_2|L\rangle_4]\\
+&[|H\rangle_1|V\rangle_3-|V\rangle_1|H\rangle_3]|+\rangle[|L\rangle_2|R\rangle_4-|R\rangle_2|L\rangle_4]\}.
\end{split}
\label{swap2}
\end{equation}
After the photon polarization and the spin measurements, photons 2 and 4 get entangled in the four Bell states, which are unequivocally
associated to the measurement results consisting of polarizations of photons 1 and 3 in the $|H\rangle$, $|V\rangle$ basis and the
spin in the $|+\rangle$, $|-\rangle$ basis (see Table \ref{tab2}).

As discussed in our previous work, \cite{hu08b} the spin-cavity unit can work as a photon entangler which generates
arbitrary entanglement including Bell states, GHZ states, and cluster states. Therefore, we think this spin-cavity unit is generally
an arbitrary entanglement generator and analyzer.  For instance, to analyze a N-photon GHZ state, we can cascade the Bell-state analysis
to get the full information on the state structure. To our knowledge, only a partial GHZ-state analyzer has ever been reported. \cite{pan98b}
With the GHZ-state analyzer, our scheme can be extended to quantum teleportation based on multi-particle entanglement. \cite{bose98}

\begin{table}[ht]
\caption{The correspondence between the photon 1, 3 polarization and the spin measurement results and the photon 2, 4 Bell states. }
\begin{ruledtabular}
\begin{tabular}{lll}
\textrm{Photons 1, 3}&
\textrm{Spin} &
\textrm{Photons 2, 4}\\
\colrule
$|H\rangle_1|H\rangle_3$ or $|V\rangle_1|V\rangle_3$ & $|-\rangle$ & $[|L\rangle_2|L\rangle_4-|R\rangle_2|R\rangle_4]/\sqrt{2}$ \\
$|H\rangle_1|V\rangle_3$ or $|V\rangle_1|H\rangle_3$ & $|-\rangle$ & $[|L\rangle_2|L\rangle_4+|R\rangle_2|R\rangle_4]/\sqrt{2}$ \\
$|H\rangle_1|H\rangle_3$ or $|V\rangle_1|V\rangle_3$ & $|+\rangle$ & $[|L\rangle_2|R\rangle_4+|R\rangle_2|L\rangle_4]/\sqrt{2}$ \\
$|H\rangle_1|V\rangle_3$ or $|V\rangle_1|H\rangle_3$ & $|+\rangle$ & $[|L\rangle_2|R\rangle_4-|R\rangle_2|L\rangle_4]/\sqrt{2}$ \\
\end{tabular}
\end{ruledtabular}
\label{tab2}
\end{table}

If we go further, this spin-cavity unit is already a quantum computer as only a quantum computer can generate and identify
arbitrary entanglement. The two-qubit phase shift gate  described by $\hat{U}(\pi/2)$ is universal when assisted by arbitrary single-qubit gates, \cite{barenco95a, lioyd95} so all quantum logic operations (including CNOT gate and CZ gate) can be built from it. Details will be
discussed elsewhere. \cite{hu10}

\section{Bell-state Analyzer (Type II)}

\begin{figure}[ht]
\centering
\includegraphics* [bb= 23 291 522 675, clip, width=7cm, height=5.8cm]{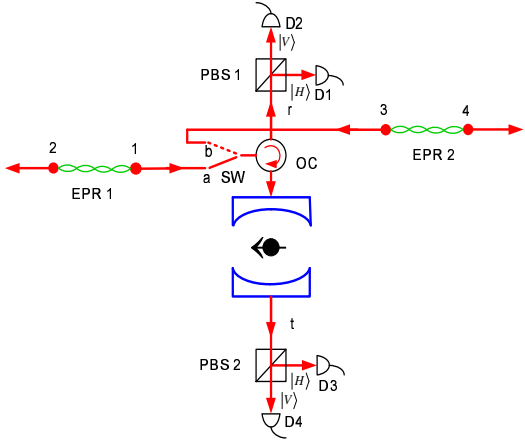}
\caption{(color online). Schematic of entanglement swapping with a QD-spin in a double-sided microcavity (type-II).  In this case,
the entanglement beam splitter works as a five-qubit GHZ-state generator and
a complete Bell-state analyzer. This unit can also be used for state teleportation (not shown here). SW (optical switch), OC (optical circulator),
PBS 1 and PBS 2 (polarizing beam splitters), D1-D4 (photon detectors). } \label{fig4}
\end{figure}

In this Section, we consider another type of spin-cavity unit (type II) with a single spin in a double-sided optical microcavity where the top and
bottom mirrors are both partially reflective. In this spin-cavity system, GCR manifests as the different reflection/transmission coefficients
between $|R\rangle$ and $|L\rangle$ photons. This allows us to make another
photon-spin entangling gate - entanglement beam splitter. \cite{hu09}
The transmission and reflection operators for this entanglement beam splitter are defined as
\begin{equation}
\begin{split}
&\hat{t}(\omega)=t_0(\omega)(|R\rangle\langle R|\otimes |\uparrow \rangle\langle \uparrow|+|L\rangle\langle L|\otimes
|\downarrow \rangle\langle \downarrow|)\\
&\hat{r}(\omega)=r_h(\omega)(|R\rangle\langle R|\otimes |\downarrow \rangle\langle \downarrow|+|L\rangle\langle L|\otimes
|\uparrow \rangle\langle \uparrow|),
\end{split} \label{ebs}
\end{equation}
where $t_0(\omega)$ is the transmission coefficient of the empty (cold) cavity, and $r_h(\omega)$ is
the reflection coefficient of the hot cavity. If $\omega=\omega_0$ is set, and $\kappa_s\ll \kappa$ and
$\text{g}^2\gg \kappa\gamma/2$ (e.g., the Purcell regime or the strong coupling regime)
are met , we get $t_0(\omega_0)\simeq -1$ and $r_h(\omega_0)\simeq 1$ (Ref. \onlinecite{hu09}).
By applying $\hat{r}(\omega_0)$, $\hat{t}(\omega_0)$ to the four Bell states, we find $|\Psi^{\pm}\rangle|+\rangle$ is transformed to
\begin{equation}
-[|R\rangle_1^{t}|L\rangle_2^{r}\pm|L\rangle_1^{r}|R\rangle_2^{t}]|\uparrow\rangle-
[|R\rangle_1^{r}|L\rangle_2^{t}\pm|L\rangle_1^{t}|R\rangle_2^{r}]|\downarrow\rangle
\end{equation}
with one photon reflected and another transmitted, and $|\Phi^{\pm}\rangle|+\rangle$ is transformed to
\begin{equation}
[|R\rangle_1^{t}|R\rangle_2^{t}\pm|L\rangle_1^{r}|L\rangle_2^{r}]|\uparrow\rangle+
[|R\rangle_1^{r}|R\rangle_2^{r}\pm|L\rangle_1^{t}|L\rangle_2^{t}]|\downarrow\rangle
\end{equation}
with two photons both reflected or both transmitted.

We can distinguish $|\Psi^{\pm}\rangle$ from  $|\Phi^{\pm}\rangle$ by simply discriminating the two photons in the same or different output
ports, rather than the spin measurements in the type-I unit. The photon polarization measurement in the $|H\rangle$, $|V\rangle$
basis can distinguish between $|\Psi^{+}\rangle$ and $|\Psi^{-}\rangle$ and between $|\Phi^{+}\rangle$ and $|\Phi^{-}\rangle$.
Obviously, the type-II spin-cavity unit is also a complete BSA with a built-in spin memory
and can be used for either state teleportation
or entanglement swapping (see Fig. 3).  Generally, this unit can also generate and analyze arbitrary entanglement as the photon-spin
entangling gate described by Eq. (\ref{ebs}) is universal together with single-qubit gates. \cite{hu09, hu10} Recently, Bonato et al constructed
a complete BSA by combining the entanglement beam splitter with an external interferometer. \cite{bonato10} In their configuration, the
spin measurement (rather than the port discrimination) is used to distinguish $|\Psi^{\pm}\rangle$ from  $|\Phi^{\pm}\rangle$.
With no external interferometer included, we think our design is stable and simple.

For type-II BSA, the fidelity and efficiency  is similar to that for entanglement generation which has been
discussed in our previous work. \cite{hu09} Both type-I and type-II BSAs can work in the weak and strong coupling regime, but high fidelity and high efficiency can only be achieved when the side leakage and cavity loss is low.

As $t_0(\omega_0)=-r_h(\omega_0)=-1$, there is a phase shift of $\pi$ between the transmitted and reflected states,
which is also recognized by Waks and Vu\v{c}kovi\'{c}. \cite{waks06} As a result,
a single photon in the $|R\rangle$
or $|L\rangle$ state can also induce the $180^{\circ}$ spin rotation around the optical axis
in the type-II structure,
which could be applied in spin echoes to prolong the spin coherence as in
the type-I structure (see Sec. III).

\section{Conclusions}

We have developed a scheme for efficient, heralded and loss-resistant quantum teleportation by exploiting the coherent
photon-spin interaction in a spin-cavity QED system. In the ideal case, this scheme is deterministic, but with reduced efficiency when losses are included. State teleportation and entanglement swapping is heralded by the sequential
detection (rather than the coincidence measurement) of two photons at different arrival times ($\Delta t<T^e_2$), and is finished
after the spin measurement. In this scheme, the spin-cavity unit works as a GHZ-state generator and a complete
Bell-state analyzer with a built-in spin memory, but generally it is  an arbitrary entanglement generator and analyzer (i.e.,
a quantum computer). The scheme can thus be extended to teleportation based on multi-particle entanglement.
As the spin-cavity unit provides a photon-spin interface, \cite{hu08b, hu09} efficient teleportation between separated
spin qubits in cavities is also possible.

We have shown that these schemes could be realized with current technology.
In the weak coupling regime, unity fidelity and $<10\%$ efficiency could be achieved.
In the strong coupling regime, we expect
near unity fidelity and $>35\%$ efficiency ($100\%$ efficiency in the ideal case).

We have also introduced a scheme for optical spin manipulations with the photon-spin entangling gates, which
could be applied for extending the spin coherence time via the spin echo techniques.
The versatile spin-cavity systems can be applied in all aspects of
quantum information science and technology, not only for  large-scale quantum communication networks, but also for
 scalable quantum computing with either photons or spins as qubits.

\section*{ACKNOWLEDGMENTS}

We thank M. Atat\"{u}re, S. Bose, S. Popescu and J.L. O'Brien for helpful discussions. This work is partly funded by
QAP (Contract No. EU IST015848) and ERC advanced grant QUOWSS.


\begin{references}
\bibitem{rate}We assume $\sim 5~$ps photon pulses and $\sim 100~$GHz repetition rate to achieve maximum enhancement.
\bibitem{note0}The reflection operator $\hat{r}$ defined by  Eq. (\ref{gate2}) is rigorous and
does not require the weak excitation approximation. Any approximation just affects the expressions
of $r_0(\omega)$ and $r_h(\omega)$.
\bibitem{note1}The solutions of the equation $\varphi_h(\omega)-\varphi_0(\omega)=\pm \pi/2$ depend
on the values of $\text{g}/(\kappa+\kappa_s)$ and $\kappa_s/\kappa$, and are two-fold degenerate with the
positive and negative frequency detunings. For $\kappa_s/\kappa<1.0$, there are
six solutions at most; For $1.0\leq \kappa_s/\kappa \leq1.3 $, there are four solutions at most; For  $\kappa_s/\kappa >1.3$,
there is no solution. In our previous work (Refs. \onlinecite{hu08a} and \onlinecite{hu08b}), we discussed two solutions
which lie closest to the central frequency in the strong coupling limit with $\text{g}/(\kappa+\kappa_s) \geq 1.5$,
and yield the highest fidelity and efficiency.
In Fig. 2, we present all solutions in both the strong and the
weak coupling regime.
\bibitem{note2}In the laboratory, we have recently achieved $\text{g}/(\kappa_s+\kappa)\geq 1.2$ for
a $d=2.5~\mu$m micropillar microcavity containing self-assembled In(Ga)As/GaAs QDs, and
have seen phase shifts up to $0.1$ radian limited primarily by the side leakage and cavity loss.
\bibitem{note3}$\text{g}$ and $\kappa$ can be controlled independently as $\text{g}$ is determined by the
trion oscillator strength and the cavity modal volume, whereas $\kappa$ by the cavity quality factor only.
\bibitem{note4}By taking $\text{g}/(\kappa+\kappa_s)=1.0$, $\kappa_s/\kappa=0.7$ and $\gamma/\kappa=0.1$ which
are experimentally achievable for a $d=1.5~\mu$m micropillar microcavity with $Q=1.7\times10^4$ (see the text),
we get the critical photon number $n_0=2\times 10^{-3}$ and the cavity lifetime $\tau=9~$ps. Therefore
the time interval between two photons should be longer than $\tau/n_0=4.5~$ns to
make the weak excitation approximation valid. Note that the photon coherence time
should be longer than the cavity lifetime so that the phase shift can be well defined (Ref. \onlinecite{hu08b}).
\end{references}
\end{document}